# Spectral Correlation Measurements at the Hong-Ou-Mandel Interference Dip*

T. Gerrits[1], F. Marsili[2], V. B. Verma[1], L. K. Shalm[1], M. Shaw[2], R. P. Mirin[1], and S. W. Nam[1]

[1]*National Institute of Standards and Technology, 325 Broadway, CO 80305, USA*
[2]*Jet Propulsion Laboratory, 4800 Oak Grove Dr., Pasadena, California 91109, USA*

thomas.gerrits@nist.gov

**Abstract**

We present an efficient tool capable of measuring the spectral correlations between photons emerging from a Hong-Ou-Mandel interferometer. We show that for our spectrally factorizable spontaneous downconversion source the Hong-Ou-Mandel interference visibility decreases as the photons' frequency spread is increased to a maximum of 165 nm. Unfiltered, we obtained a visibility of 92.0 ± 0.2 %. The maximum visibility was 97 ± 0.2 % after applying filtering. We show that the tool can be useful for the study of spectral correlations that impair high-visibility and high-fidelity multi-source interference applications.

**Letter**

Multi-photon entanglement lies at the heart of many quantum information and communication applications [1-8]. To generate the photonic states needed for these systems in a scalable manner it is often necessary to entangle photons from different sources together. This can be accomplished by interfering two photons with one another, but the process is only successful if both photons are completely indistinguishable from one another in every degree of freedom. Such indistinguishable photons are also critical to implementations of boson sampling and quantum random walk protocols which rely on quantum interference with a high visibility [9-13]. Generating indistinguishable photons from two different sources remains a challenging open problem [14-19]. One of the most common methods of generating entangled photons is through the process of spontaneous parametric downconversion where a parent photon splits into a pair of daughter photons, called the signal and idler. The photon pairs generated through downconversion can possess strong spectral entanglement and correlations. When interfering two photons from different downconversion sources, these correlations can lead to distinguishing spectral information that spoils the interference. It is therefore important to engineer the downconversion process to produce photon pairs that are spectrally uncorrelated with one another [14]. To date, much effort has been spent developing ways to design spectrally pure downconversion sources by studying different material properties, angles of emission of the photons [20, 21], and the nonlinear profile of the crystal [22, 23].

A standard method for measuring how distinguishable two photons are is to use a Hong-Ou-Mandel interferometer (HOM) [19]. In a typical HOM setup two photons interact on a beamsplitter. If the photons are indistinguishable, they will interfere and both leave the same port of the beam splitter. If any distinguishing information is present, there is a chance that the two photons will exit from different beamsplitter ports. By monitoring the output ports of the beamsplitter for coincidence events and changing the relative time of arrival of the two photons on the beamsplitter, it is possible to measure the amount of distinguishing information present.

However, a standard HOM measurement only reveals the amount of distinguishing information, but does not provide clues as to its origin. Here we present a new tool for studying the two-photon interference and indistinguishability of two photons that is well suited to analyzing the spectral properties of photons. We take two photons and interfere them on a beamsplitter in a Hong-Ou-Mandel interferometer [24]. Using a fiber-based spectrometer we measure joint-spectral information from the two photons as a function of their arrival time on the HOM interferometer [18]. This provides detailed dynamical information about the spectral correlations not typically available in a HOM measurement. This extra information yields direct insights into the quality and behavior of the photon source. Our technique also enables us to implement arbitrary spectral filters to shape and tailor the measured joint spectrum of the two photons.

We use a single downconversion source for the demonstration of the spectrally resolved Hong-Ou-Mandel interference. The spectral intensity distribution of the two-photons, $I(\omega_1,\omega_2)$, exiting from opposite ports of the HOM interferometer is given by [25, 26]:

$$I(\omega_1,\omega_2) \propto \frac{1}{2}\left|C(\omega_1,\omega_2)e^{i(\omega_1 t_1+\omega_2 t_2)} - C(\omega_2,\omega_1)e^{i(\omega_2 t_1+\omega_1 t_2)}\right|^2, \quad [1]$$

where $C(\omega_1,\omega_2)$ is the complex joint spectral amplitude (JSA) of the two-photon wavefunction, $\omega_1$ and $\omega_2$ are the signal and idler frequencies respectively, and $t_1$ and $t_2$ are the signal and idler photon detection times. The beamsplitter acts as a symmetry filter. If the joint spectrum $C(\omega_1,\omega_2)$ is symmetric under the exchange of the signal and idler frequencies, then when $t_1 = t_2$ perfect destructive interference takes place and the two photons never exit from opposite ports of the beamsplitter. A typical photon generated from downconversion comes in a wavepacket containing many different frequency components. Interference of such photons leads to a rich interference spectrum that beats as the product between the relative arrival time, $\delta t = t_1 - t_2$ and the frequency difference, $\Delta\omega = \omega_1 - \omega_2$.

Typical single-photon detection systems integrate over all of the frequency components, leading to a coincidence detection rate of:

$$R_c = \iint d\omega_1 d\omega_2 I(\omega_1,\omega_2). \quad [2]$$

Such a measurement ignores the additional information and complexity present in the spectrum of interference between the individual frequency components of Eq. 1. If instead the full interference spectrum, $I(\omega_1,\omega_2)$, can be directly measured, then deeper insight into the dynamics of the HOM dip can be obtained. For sources where the two photons are not indistinguishable, it is possible to determine exactly which frequency components in the photon wavepackets contribute to the distinguishing information.

A pair of daughter photons generated from downconversion can share strong spectral correlations. To achieve good interference visibility between the signal and idler photons from a single downconversion event it is necessary to engineer the JSA so that it is symmetric around the bi-photon's zero detuning axis ($\omega_1 - \omega_2 = 0$) and $C(\omega_1,\omega_2) = C(\omega_2,\omega_1)$ [25, 26].

When interfering two photons from different downconversion events, the spectral correlations between the signal and idler of each pair lead to distinguishing information. Therefore, it is necessary to design a source that eliminates these spectral correlations. In this case the JSA of each photon pair needs to be factorizable, meaning that $C(\omega_1,\omega_2) = \phi(\omega_1)\theta(\omega_2)$, in order to obtain high visibility interference.

Here, we demonstrate a HOM interferometer that uses a fiber-based spectrometer capable of simultaneously recording all of the spectral correlations that exist. This enables us to directly observe the frequency dependent interference between two photons at different positions in the HOM dip. To measure the two-photon frequency correlations we send each photon through a dispersive medium, long single-mode optical fibers in our case, and then measure the arrival time with fast single-photon detectors. The dispersive medium couples each photon's frequency with its arrival time. By accurately recording the arrival times of the photons it is then possible to determine the frequency correlations between the pair. From this information it is possible to reconstruct the joint-spectral intensity (JSI) distribution of the two photons at different points in the HOM dip.

Our pair of photons is generated using spontaneous parametric downconversion (SPDC) in a $KTiOPO_4$ (pp-KTP) crystal with a poling period of 46.15 μm [18]. The source is engineered to produce nearly pure, degenerate single photons in orthogonal polarization modes at a center wavelength of 1570 nm, and is pumped with a femtosecond-mode Ti:Sapphire laser at a repetition rate of 76 MHz and a center wavelength of 785 nm (see Figure 1a). The spectral full-width half maximum (FWHM) of the Fourier transform-limited pump laser pulse is 5.3 nm, and its temporal FWHM is 140 fs. We focus the pump beam to a beam waist of 57 μm FWHM using a 100 mm focusing lens in front of the pp-KTP crystal. The emerging two-mode state is collimated through a 100 mm collimation lens and sent to a polarizing beam splitter (PBS). After passing through an optical path delay, the two modes are recombined at a second polarizing beamsplitter. The arrival-time of idler photon can be adjusted by moving a translation stage (Δt), allowing the scan of the HOM interference dip. The combination of a half-wave-plate (HWP) and a third PBS serves as a variable transmission/reflection beamsplitter. However, in this study we only use a splitting ratio of 50/50 needed for the HOM interference measurement, and a 0/100 splitting ratio for measuring the JSI of the source. The spectral output FWHM bandwidth of the source is 17.3 nm. We checked the HWP-PBS combination for wavelength dependent splitting ratios, as these could cause the HOM interference to degrade. We found that the splitting ratio varied from 0.48 at 1640 nm to 0.51 at 1510 nm. This non-ideal splitting ratio has less than a 1% effect on the measured HOM interference visibility[1] as it only dominates for low photon pair probabilities in the wings of the SPDC.

Photons emerging from the two output ports of the beam splitter are coupled into single-mode fibers using achromatic lenses with a focal length of 15.4 mm. We estimate the combined transmission through the free-space optics and coupling efficiency into fiber to be 42 ± 1 % and 47 ± 1 % for the vertically-polarized (V) signal and horizontally-polarized (H) idler photons respectively using the Klyshko method [27][2]. We use 1.3 km and 2.3 km lengths of single-mode fiber, which have an optical transmission of 77 % and 87 % respectively at 1550 nm, as the dispersive medium in our fiber spectrometer. After travelling through the fibers,

---

[1] Imperfect splitting ratio has to be addressed and should pose a concern when seeking perfect-visibility interference from a broadband source

[2] At low pair production rates (<0.5%), we determined the ratio between coincidences and singles at a 0/100 HWP-PBS splitting ratio for signal and idler paths yielding the overall system efficiency including the detectors

the photons are sent to high-efficiency, low-jitter, superconducting nanowire single photon detectors (SNSPDs) [28]. The time-of-arrival of each photon is then recorded with commercial time-stamping electronics.

Four detectors, labeled channels 1-4 in Figure 1a, were used with detection efficiencies ranging from 67 % to 87 % at 1550 nm, respectively. The timing jitter varied from 120 ps to 175 ps for these four detectors and, along with the fiber dispersion, determines the resolution of our fiber spectrometer. A full description of the SNSPDs and the fiber spectrometer is given in the supplemental material. The SPDC source was single pass with a timing uncertainty of less than one picosecond - much less than the jitter in the detectors which limits the resolution of the spectrometer. Near 1570 nm, the dispersion of the 1.3 km (2.3 km) length was 24.0 ps/nm (41.9 ps/nm). Calibration of the optical fibers, shown in Fig. 1c, was performed following the method described in Ref. [18]. Combined with the jitter of our SNSPDs (see Fig. 1b) this results in a finite spectral resolution for detecting a single photon and coincidence detection. The spectral resolutions are summarized in table I. The spectral resolution of the fiber spectrometer in the coincidence basis is given by the convolution of both single-photon instrument response functions for each of the SNSPDs. In principle, the resolution can be enhanced by use of lower jitter detectors [29] and by increasing the optical fiber length. However, increasing the optical fiber length will add significant loss, and the uncertainty in the time-of-arrival will depend on temperature fluctuations in the fibers. We integrated each spectral correlation matrix for 10 minutes when measuring the frequency resolved HOM interference.

| Channel | Detector Efficiency | Detector Jitter FWHM (ps) | Single photon spectral resolution FWHM (nm) | Channel Combination | Coincidence detection spectral resolution FWHM (nm) |
|---|---|---|---|---|---|
| Ch1 | 87 ± 1 % | 120 | 5.0 | Ch1 & Ch2 | 3.9 |
| Ch2 | 85 ± 1 % | 150 | 6.3 | Ch3 & Ch4 | 2.7 |
| Ch3 | 67 ± 1 % | 175 | 4.2 | Ch1 & Ch3 | 3.2 |
| Ch4 | 81 ± 1 % | 150 | 3.6 | | |

Table I: Summary of spectral resolution, detector jitter and detection efficiency at 1550 nm.

Figure 2(a) shows the raw JSI showing the photons' time-of-flight correlations. The axes represent the individual arrival times for a coincidence between signal and idler photons. The slight curvature to the spectrum is due to the wavelength-dependent dispersion of the fiber. The horizontal and vertical axes show the arrival times for photons that traveled through the 1.3 km and 2.3 km lengths of fiber, respectively. After correcting for the wavelength-dependent dispersion of the individual fibers, the resulting |JSA| is shown in Fig. 2(c). The log-scale plot of the |JSA| shown in Fig. 2(b) clearly shows the sinc-lobes associated with the phase-matching condition of our source. From the measured JSI we can compute the spectral purity, or factorizability, of our source using the Schmidt decomposition [30]. A Schmidt number of 1 corresponds to a completely factorizable source while a large Schmidt number implies a high degree of correlation. We measure a Schmidt number of 1.07, in agreement with results reported in Ref. [18]. Note that the total measured bandwidth of the output of our source including the sinc-lobes ranges over 20 THz (165 nm).

The central plot of Fig. 3 shows the measured HOM interference dip and the bunched two-state component emerging from either port of the HOM beamsplitter. The insets show snapshots of the measurement, theory, predicted shape and the spectral correlations of the bi-photons after interference at the HOM beam splitter at various HOM delays. The complete set of data is presented in an animation accompanying this paper as supplementary material. The blue crosses of the central plot show the integral of each spectral correlation, equivalent to the conventional integral coincidence-counting method. The visibility of the measured HOM interference dip is 92.0 % ± 0.2 %. Note that all data presented in this paper are unfiltered and no background subtraction was applied.

The dotted black line in the central figure shows the theoretically predicted HOM interference based on Eq. 1 and a first-principle calculation using the Sellmeier parameters for our crystal [31]. This calculation takes into account the full phase dependence of the JSA. Based on the measured JSI, it is also possible to directly place an upper bound on the HOM interference visibility. As we only possess information about the intensity of the joint spectrum and not its phase we must make the additional assumption that the joint spectrum is real. In this case Eq. 1 can be rewritten as $I(\omega_1, \omega_2) \propto \frac{1}{2}\left[C^2 + C_T^2 - CC_T e^{i\Delta\omega\delta t} - C_T C e^{-i\Delta\omega\delta t}\right]$, where $C = C(\omega_1, \omega_2)$ and $C_T = C(\omega_2, \omega_1)$, and are both real. This will provide an estimate of the maximum dip visibility possible – any complex phase terms between different frequency components will only serve to lower the dip visibility. The red line in the central figure shows the maximum HOM dip possible directly calculated in this way from the measured |JSA| in Fig. 2(c). Both calculations represent the measured HOM dip well. However, the dip visibility is not recovered in the data as predicted by the calculations. The predicted visibilities are 98.3 % and 99.5 % for the red and dotted black line, respectively.

Based on Eq. 1, the visibility should reach a value close to 100% if the JSA is symmetric with respect to the bi-photon's zero detuning axis. In fact, our JSA is not perfectly symmetric around the zero-detuning axis, as our focusing and collimation setup slightly biased the JSA towards larger signal and smaller idler detuning of about 0.1 THz [31]. As we will show below, the main source for decreased visibility lays in the sinc-lobes of the JSA, and we believe that spatial mode distinguishability originating from slight misalignments is the cause for some of the reduced visibility. The bunching of the two-photon state emerging from either beamsplitter output port is shown by the green and black solid lines in the central plot. Coincidences between detector channels 1 (3) and 2 (4) detect the presence of two photons in beam splitter output port 1 (2). The temporal shape of the two-photon bunching resembles the shape of the HOM dip, as it is to be expected.

Four different spectra are shown in each inset of Fig. 3 at various HOM delays: The measured spectral correlations between the two output ports of the HOM beam splitter (lower left figure); The predicted spectral correlations between the two output ports of the HOM beam splitter based on the simplified Eq. 1 and the measured |JSA| (lower right figure); The measured spectral correlations of the two-photon state, *i.e.* bunching, in output port 1 (2) of the HOM beam splitter - upper left (right) figure.

All insets show the predicted interference as predicted by Eq. 1. The period of the interference is proportional to $\Delta\omega\delta t$, linear with spectral detuning from the center frequency, symmetric around $\Delta\omega = 0$ and linear with HOM interference delay. High frequency interference patterns are visible at large $\delta t$ in the theoretical prediction. However, due to the spectral resolution of the fiber spectrometer, the measurements show low frequency interference patterns at small $\delta t$ only.

Using the fiber spectrometer we can post-select frequency components of the spectral correlations, and in the following we apply post-selective spectral filtering to investigate the origin of the reduced HOM interference visibility in the measurement. The black solid line in

Fig. 4(a) shows the HOM interference visibility as a function of post-selective filter bandwidth. We chose a virtual top hat filter around the center of the HOM spectral intensity distributions and calculated the expected HOM dip visibility by integrating the portion of the photons passing through this filter. Certainly, many other filter shapes (even those not feasible using optical components) can be constructed based on this approach. Figure 4(a) shows the HOM interference visibility decreasing as the filter bandwidth increases. The maximum visibility for a filter bandwidth equivalent to the resolution of the fiber spectrometer (0.44 THz) is 97%. The reduced visibility at minimum filter bandwidth is most likely caused by a number of small experimental imperfections, such as: dark counts, fluorescence and double pair generation, misalignment, wavefront distortion, non-perfect 50/50 beamsplitting for the center frequencies, and detector jitter causing distinguishable frequency components further away from the center to spoil the maximum visibility. The visibility drops to about 92 % for the maximum filter bandwidth of 15 THz. Beyond about 5 THz bandwidth photons only exist in the sinc-lobes of the JSA. At large detunings from the center wavelength the spatial distributions of the pairs are quite different, leading to distinguishability in the HOM interference. In our setup though, because we use a single-mode fiber, the spatial modes should be matched based on the overlap of the fiber mode with the spatial mode of the photons. If, however, the spatial overlap at the HOM beam splitter is different from the mode collected by the optical fiber, the photons are distinguishable. Figure 4(b) shows such a scenario. We assume two modes exiting from the two output ports of the 50/50 HOM beam splitter. The collection from output port 1 is aligned perfectly with the optical axis and the light collected from output port 2 has a slight misalignment with the optical axis ($\theta$). Figure 4(c) shows the HOM interference visibility as a function of the misalignment angle $\theta$. A clear drop in the visibility is observed for angles smaller than 1°. The red line in Fig. 4(a) shows the anticipated HOM dip visibility as a function of virtual filter bandwidth for a misalignment angle of 0.5°. The predicted values for the HOM visibility are reduced by 3% to compensate for the experimental imperfections mentioned above. The predictions qualitatively follow the trend of the observed visibility decrease. However, the exact shape of the measure data cannot be predicted fully.

In the near future, we plan to further study the limits imposed by imperfect spatial mode overlap by implementing a fiber-based HOM beam splitter ensuring that the spatial modes are well matched for each photon pair coupled into the optical fiber.

**Conclusion**

We presented a novel tool capable of efficiently measuring joint spectral correlations emerging from the two output modes of a HOM interference setup. We show that fast acquisition of the correlation spectra is possible, and that the reduced HOM interference visibility is mostly due to pairs far detuned from the center wavelength. We show that distinguishable spatial modes can be the cause of the reduced HOM interference visibility.

**Acknowledgements**

This work was supported by the Quantum Information Science Initiative (QISI). T.G. thanks A. Fedrizzi, M. J. Stevens, A. White and F. Wong for discussions during the preparation of this manuscript.

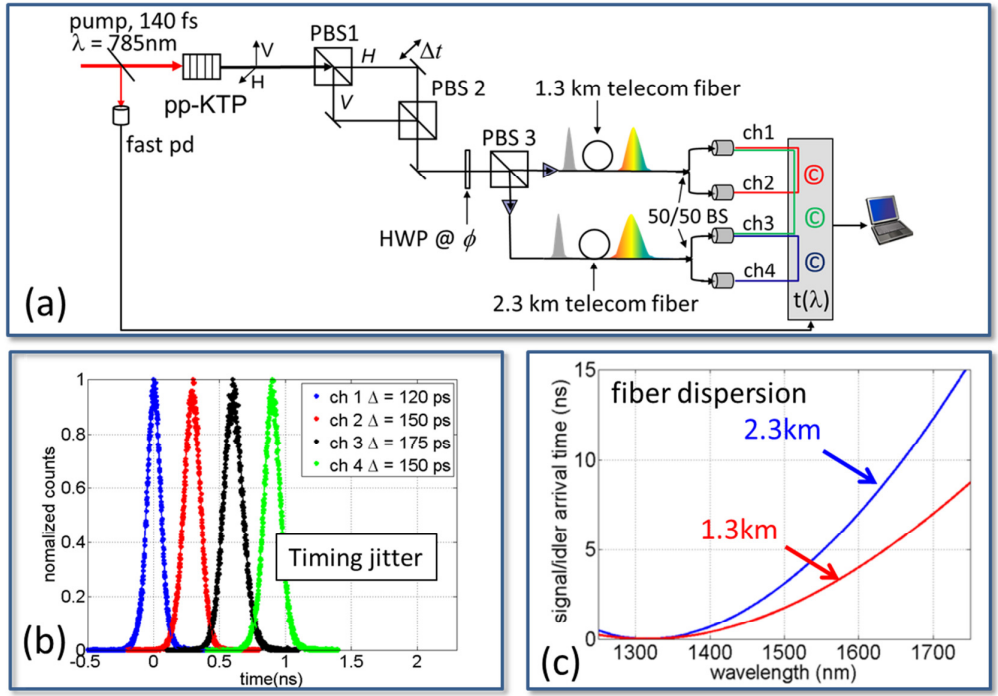

FIG. 1 (a) Experimental setup: A femtosecond pulse, pumps a pp-KTP crystal producing type-II SPDC. PBS1 and PBS2 are sandwiched around an optical path delay $\Delta t$ for the $H$ (idler) photon. A HWP at angle $\phi$ and PBS3 serve as a variable beamsplitter for both input modes. Two long single-mode fibers are used to encode the photons' frequency into time-of-arrival. Four SNSPDs probe various spectral correlations of the SPDC after the variable beam splitter. Time-stamping electronics record the arrival time of each photon; fast pd: fast photo diode (trigger). (b) Instrument response functions of all four SNSPDs (c) Measured fiber dispersion curves for both single-mode fibers.

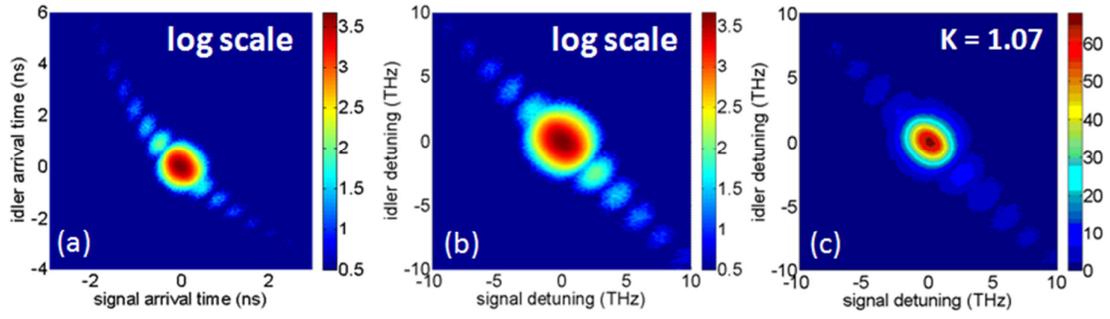

FIG. 2 Joint probability distribution, integrated for two minutes (a) As measured, log-scale plot of the raw joint time-of-arrival probability distribution (b) Log-scale plot of the joint spectral intensity (JSI) (c) Absolute value of the joint spectral amplitude (JSA) distribution.

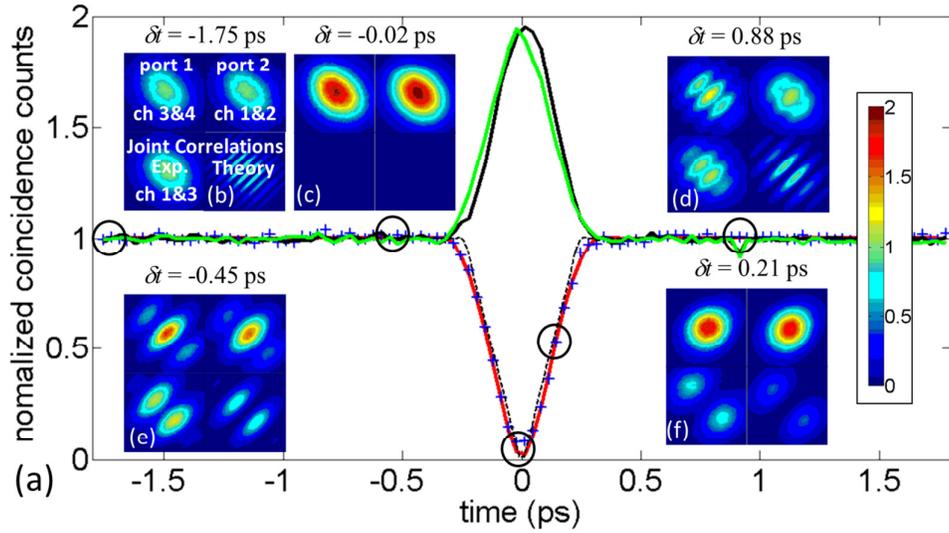

FIG. 3 (a) HOM interference dip; blue crosses: Integral of the correlation spectra as a function of HOM delay; black dotted line: first-principle calculation of the HOM interference dip based on Eq. 1; red line: HOM interference dip prediction based on the simplified version of Eq. 1 and the measured |JSA|; green and solid black line: bunching of two-photon component at either output port of the beamsplitter; (b)-(f) correlation spectra at various HOM delays; each inset consists of four spectra. The scale is ± 2THz for both axes; upper left (right): two-photon component, *i.e.* bunching, in output mode 1 (2), lower left: measured spectral correlation of HOM interference, lower right: spectral correlation of HOM interference based on Eq. 1 and the measured |JSA|. Bracket-insets denote the detector channel numbers used for the measurement.

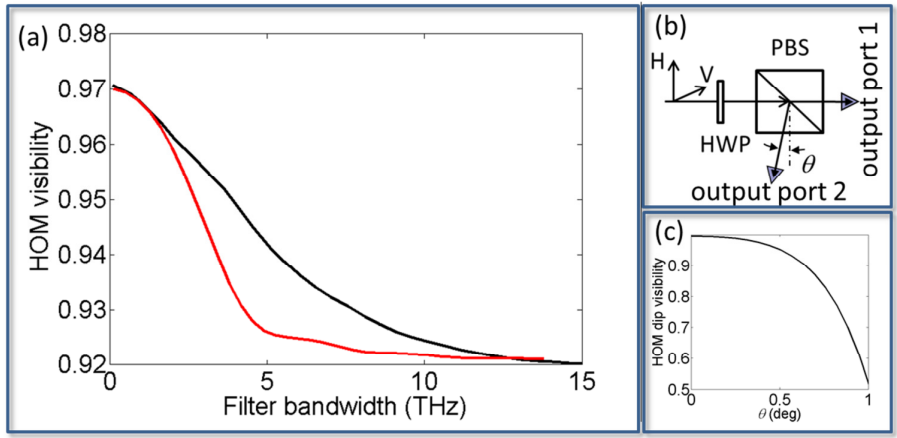

FIG. 4 (a) Red line: Measured HOM dip visibility versus virtual filter bandwidth, based on a variable bandwidth top-hat post-processing filtering method. Red line: First principle calculation assuming asymmetric mode fiber coupling; (b) schematic of the asymmetric mode coupling; (c) calculation of maximum HOM visibility as function of off-optical-axis collection angle, $\theta$

# Supplemental Material
## Spectral Correlation Measurements at the Hong-Ou-Mandel Interference Dip

T. Gerrits[1], F. Marsili[1], V. B. Verma[1], L. K. Shalm[1], M. Shaw[2], R. P. Mirin[1], and S. W. Nam[1]

[1]National Institute of Standards and Technology, 325 Broadway, CO 80305, USA
[2]Jet Propulsion Laboratory, 4800 Oak Grove Dr., Pasadena, California 91109, USA

The Supplemental Material consists of a detailed description of the superconducting nanowire single photon detectors (SNSPDs) used in the experiment and a detailed description of the fiber spectrometer capabilities.

**Superconducting nanowire single photon detectors**

Superconducting nanowire single photons detectors (SNSPDs) [1] are based on thin (thickness below 10 nm) and narrow (width below 200 nm) superconducting wires current-biased close to their superconducting critical current. The absorption of one or more photons in a superconducting nanowire drives part of the nanowire to the normal state with a resistance on the order of several k$\Omega$ [2-4], which can be detected by using an appropriate readout circuit [5]. Our detector system employed four SNSPDs based on amorphous tungsten-silicide ($W_xSi_{1-x}$, or WSi) nanowires [6]. The system detection efficiencies at 1550nm were 87 ± 1 %, 85 ± 1 %, 67 ± 1 %, and 81 ± 1 % for channels 1, 2, 3 and 4, respectively. All channels had a < 1 s$^{-1}$ intrinsic dark-count rate, and background-limited count rate (*BCR*) of ~ 300 s$^{-1}$. The SNSPDs had an active area of 15 µm × 15 µm and were made from 110 nm wide nanowires. The devices were fabricated on Si substrates and embedded in an optical stack to enhance the detector absorption. The optical stack was designed for allow front-side illumination, and was composed of the following layers from top (illumination side) to bottom: $TiO_2$, $SiO_2$, WSi nanowires, $SiO_2$, and Au. We used silicon micro-machining to implement a self-aligned packaging scheme with telecom single-mode optical fibers as described in Ref. [7]. This scheme allows efficient coupling between the core of a single-mode optical fiber and the active area of the detector with ± 3 µm precision. The devices could be operated in the temperature range 150 mK to 1 K without degradation of the detector performance.

**Fiber Spectrometer**

A dispersive medium, such as a single-mode optical fiber, allows encoding of the frequency of a photon with known creation time into time-of-arrival at the exit of that medium. During a parametric downconversion event, two photons are born at the same moment. This moment is known with an uncertainty equal to the optical length of the non-linear medium in which the photon pair was created. For single-pass SPDC sources this uncertainty is small and generally in the picosecond regime. This is in contrast to when SPDC is produced, using a resonant cavity. In that case, the emerging pairs will have a temporal uncertainty given by the cavity lifetime, which can be many orders of magnitude larger than for single-pass SPDC. In addition, the timing jitter of the single-photon detector adds to the photon's arrival time uncertainty. For a fiber spectrometer, the dispersion imposed by the optical fiber must be larger than the combination of all of these other uncertainties. The ratio of photon-arrival uncertainty and fiber dispersion determines the resolution of the fiber spectrometer

**Coincidence based measurements with fiber spectrometer**

Each coincidence count between the SNSPDs delivers one datapoint in the joint spectral probability distribution, and is therefore much more efficient than a two-dimensional scanning approach, *i.e.* two monochromators each followed by a single photon detector. This, along with high system detection efficiencies leads to much shorter acquisition times for a joint spectral probability distribution than is possible with the two-dimensional scanning approach. The rate at which coincidences are detected is given by : $R_s = R_p \eta_1 \eta_2$, where $R_p$ is the pair production rate, $\eta_1$ and $\eta_2$ are the two system detection efficiencies of both modes from the source including the detector. To illustrate, we calculate examples for a given parameter set. Assuming a pair generation probability of 0.1% per pump pulse, a laser repetition rate of 76 MHz, an overall system efficiency of 30% for both modes, the resulting coincidence rate then is about 7 kHz.

We can now estimate the uncertainty per bin in the coincidence matrix from the counting statistics as a function of integration time and coincidence rate. Assuming 100 bins of the coincidence matrix are within the FWHM of the joint spectral probability distribution, we calculate statistical uncertainties of 5%, 1.6% or 0.5% when integrating for 5.7s, 57s or 570s. The coincidence rate is reduced by a factor of 2 when performing the spectrally resolved HOM interference experiment owing to the 50/50 HOM-beamsplitter.